\documentclass[aps,prl,twocolumn,showpacs,superscriptaddress,groupedaddress]{revtex4}  
\usepackage{amssymb,amsmath,bm,graphicx}

\begin{document}

\title{Microcanonical particlization with local conservation laws}

\author{Dmytro Oliinychenko}
\author{Volker Koch}

\affiliation{Lawrence Berkeley National Laboratory, 1 Cyclotron Rd, Berkeley, CA 94720, US}

\begin{abstract}
 We present a sampling method for the transition from relativistic hydrodynamics to particle transport, commonly referred to as particlization, which preserves the local conservation of energy, momentum, baryon number, strangeness, and electric charge microcanonically, i.e. in every sample. The proposed method is essential for studying fluctuations and correlations by means of stochastic hydrodynamics. It is also useful for studying small systems. The method is based on Metropolis sampling applied to particles within distinct patches of the switching space-time surface, where hydrodynamic and kinetic evolutions are matched.
\end{abstract}

\maketitle

One of the key goals of modern heavy ion collision experiments is to search for a phase transition between a hadron gas and a quark-gluon plasma (QGP), and to locate the corresponding critical point. The vicinity of the critical point is characterized by enhanced event-by-event fluctuations \cite{LL_T5_par146,Stephanov:2008qz}. Therefore, considerable attention is devoted to correlation and fluctuation observables, such as proton, net-proton, net-charge, and kaon cumulants~\cite{Luo:2015ewa, Adamczyk:2017wsl}, fluctuations of various  particle ratios~\cite{Abdelwahab:2014yha}, transverse momentum correlations~\cite{Adam:2019rsf}, and charge balance functions~\cite{Adamczyk:2015yga}. Since a heavy ion reaction is a dynamical process, it is essential to study these observables within a dynamical framework. A very successful dynamical treatment is a hybrid approach~\cite{Hirano:2012kj} which combines the relativistic hydrodynamic evolution of the high (energy) density QGP phase with the kinetic transport for a more dilute hadronic phase. This approach successfully reproduces bulk observables such as particle spectra and flow (see e.g.~\cite{Ryu:2017qzn}). Switching from the continuous relativistic hydrodynamics to the discrete particle transport, often referred to as ``particlization'', is usually performed on a hypersurface characterized by constant energy density, temperature, or Knudsen number \cite{Ahmad:2016ods}. In the existing relativistic models the switching only occurs in one direction: from hydrodynamics to particles, but not \textit{vice versa}. This is in contrast to non-relativistic hybrid approaches, where dynamical domain decomposition methods are routinely applied  and the switching is performed in both directions (see e.g.~\cite{Donev:2009}). 
To study correlations and fluctuations within the hybrid approach, hydrodynamics has to be either extended by stochastic terms directly~\cite{Kapusta:2011gt,Murase:2013tma,Kumar:2013twa,Nahrgang:2018afz} or coupled to a non-equilibrium field with a stochastic noise~\cite{Nahrgang:2011mg,Herold:2014zoa}. In both cases, it is essential that the particlization preserves the fluctuations generated by such models. This is a non-trivial task, which so far has not been done in the context of relativistic hydrodynamics. In the non-relativistic case, this problem is addressed in several ways~\cite{Donev:2009}. One of them is to exactly match the fluxes at the interface, which in the relativistic case corresponds to local event-by-event conservation laws, or in other words, microcanonical sampling. The standard Cooper-Frye sampling used in relativistic models (for a detailed description see e.g.~\cite{Huovinen:2012is}), on the other hand, is grand-canonical. It combines the Cooper-Frye formula for the momentum distribution in a hypersurface cell~\cite{Cooper:1974mv} with Poissonian sampling of the multiplicity distributions. Together they result in total energy, momentum, and charges fluctuating around correct means, as it is illustrated in Fig.\ 5 of~\cite{Schwarz:2017bdg}. Additionally, in the standard procedure particles in different cells are sampled independently, and thus are uncorrelated, although their velocities may still be correlated via a common flow velocity profile. The scaled variances of multiplicitites are $\omega \equiv \frac{\langle N^2 \rangle - \langle N \rangle^2}{\langle N \rangle} = 1$ by construction. In contrast, in the microcanonical case $\omega < 1$ and particles should be correlated due to the conservation laws. Moreover, event-by-event conservation laws are important not only for correlations and fluctuations. For example, in small systems they can also affect mean values. Therefore, constructing and realizing a microcanonical sampling algorithm is the purpose of this work.

Attempts to introduce (micro)canonical  particlization have been undertaken previously, but they rely on intuition-based \textit{ad hoc} modifications to the standard sampling algorithm such as: introducing local charge conservation by sampling  particle-antiparticle pairs~\cite{Bozek:2012en}, trying to satisfy conservation laws one by one in a ``mode sampling'' algorithm~\cite{Huovinen:2012is}, and rejecting particles that increase the deviation from the desired conserved charges in the SPREW algorithm~\cite{Schwarz:2017bdg}. The multiplicity distribution sampled by these algorithms is not known precisely and does not correspond to a canonical or microcanonical ensemble. The SER algorithm~\cite{Schwarz:2017bdg,Oliinychenko:2016vkg} samples the correct canonical distribution, but extending it to the microcanonical ensemble is impossible. Note that energy and momentum conservation, which distinguish the microcanonical from the canonical ensemble, influence not only $p_T$ fluctuations, but also the fluctuations of multiplicities. This is why we propose a method to conserve all charges, energy, and momentum simultaneously.

First of all, we define regions over which conservation laws should be applied; we call these regions ``patches''. There are two requirements for the patch size $b$. First, it should be comparable to the hydrodynamics scale discussed in the context of fluctuating hydrodynamics \cite{An:2019osr}, therefore it should be much larger then the mean free path in a weakly-coupled system or a thermal length in a strongly coupled system: $b \gg 1/T$, where $T$ is temperature. Second, one patch should contain many particles, therefore $b^3 n \gg 1$, where $n$ is particle density. A patch should not necessarily be much smaller than the system size. If a system is small, then conservation laws should be applied over the whole system and by definition the whole system is one patch. However, if a patch is comparable to the system size, then the applicability of hydrodynamics may be questionable. In the usual applications of hydrodynamics it is reasonable to use every computational grid cell as a patch. Indeed, even in simulations of micro- and nanofluids the number of particles per computational cell is of the order of 100~\cite{Donev:2009}.  However, in typical simulations of relativistic ion collisions the average number of particles per computational cell is of order $10^{-3} - 10^{-1}$, given a typical cell size between (0.2 fm)$^3$ and (0.5 fm)$^3$. While the strategy of sampling ``fractional'' particles is possible~\cite{Steinheimer:2017dpb}, it leads to complications in the treatment of the fluctuations in the subsequent transport evolution. Our strategy is to split the hypersurface into independent patches of similar size, and to require conservation laws in every patch. In practice this implies that a patch consists of roughly $50 - 1000$ computational cells. A typical hypersurface consists of the order of $10^6$ cells, resulting in about $10^3 - 10^4$ patches. The exact number of patches, and therefore the ``localness'' of the conservation laws, can be treated as a parameter. For a given patch, the quantities to be conserved are
\begin{eqnarray} \label{Eq:cons_laws}
\begin{pmatrix}
P_{tot}^{\mu} \\
B_{tot} \\
S_{tot} \\
Q_{tot}
\end{pmatrix} = \sum_{\substack{ \mathrm{cells}\\i}} \int
\begin{pmatrix}
p^{\mu}_i\\
B_i \\
S_i \\
Q_i
\end{pmatrix}
\frac{p^{\nu} d\sigma_{\nu}}{p^0}
\mathit{f}_i(p^{\alpha}u_{\alpha}, T, \mu_i) \frac{ g_i d^3p}{(2 \pi \hbar)^3}
\end{eqnarray}
%
%
%
where the sum runs over all hadron species $i$ including resonances, and over all cells of the patch. Here $d\sigma_{\mu}$ is a normal four-vector of the cell (see definition in~\cite{Huovinen:2012is}), $u^{\mu}$ is the collective velocity of the cell, $T$ is temperature of the cell, and $\mu_{B,S,Q}(x)$ are the chemical potentials of the cell, responsible for the conservation of baryon number $B$, strangeness $S$, and electric charge $Q$. The chemical potential of the species $i$ is defined as $\mu_i = \mu_B B_i + \mu_S S_i + \mu_Q Q_i$, while $g_i$ is the degeneracy of the species. It may seem surprising to consider a local temperature and chemical potentials in a microcanonical sampling. However, there is no contradiction here. Conservation laws are imposed only over the whole patch. Variations in energy density, quantum number densities, and collective velocities from cell to cell within a patch are allowed and characterized by local values of $T$ and $\mu$. Preserving these local variations is important to ensure a faithful description of higher order azimuthal anisotropies~\cite{Gardim:2011xv}, which otherwise would be smeared. The probability $P$ of a given particle configuration in a patch is a product of the usual Cooper-Frye formulas and global delta-functions which guarantee conservation laws over the patch:
 \begin{eqnarray} \label{Eq:sampled_distribution}
P(N, \{N_s\}^{\mathrm{species}}, \{x_i\}_{i=1}^N, \{p_i\}_{i=1}^N) = \mathcal{N} \nonumber \\
\left( \prod_s \frac{1}{N_s!} \right) \prod_{i=1}^N \frac{g_i}{(2\pi\hbar)^3} \frac{d^3p_i}{p_i^0} p_i^{\mu}d\sigma_{\mu} \, \mathit{f}_i(p^{\nu}_i u_{\nu}, T, \mu_i) \times  \nonumber \\
\delta^{(4)}(\sum_i p^{\mu} - P^{\mu}_{tot}) \, \delta_{\sum_i B_i}^{B_{tot}}
\delta_{\sum_i S_i}^{S_{tot}} \, \delta_{\sum_i Q_i}^{Q_{tot}}
\end{eqnarray}
Note that here the number of particles of each hadron species $N_s$ is not fixed, and neither is the total number of particles $N = \sum_s N_s$. Instead, both are distributed according to Eq.~(\ref{Eq:sampled_distribution}). The quantities $d\sigma_{\mu}$, $u^{\mu}$, $T$, and $\mu_{B,S,Q}$ depend on the spatial position of a particle $x_i$. The charges  $B_{tot}$, $S_{tot}$ and $Q_{tot}$ are computed using Eq.~(\ref{Eq:cons_laws}). In practice these charges are real numbers, not integers. 
To address this problem, we suggest to either round $B_{tot}$, $S_{tot}$, and $Q_{tot}$ to nearest integers or distribute the non-integer parts according to a multinomial distribution, which is guaranteed not to obfuscate total charges on the hypersurface \footnote{A detailed discussion of how to generate the patches, what their optimal size are and how to address the rounding problem will reported on in a separate detailed publication. }.

Direct sampling of the $N$-particle probability distribution expressed by Eq.~(\ref{Eq:sampled_distribution}) is difficult due to the unknown normalization factor $\mathcal{N}$ and the $\delta$-functions. To sample it we apply a Metropolis algorithm, a Markov chain Monte Carlo method, which in our case is closely related to solving the Boltzmann equation with the stochastic rate method~\cite{Seifert:2017oyb}. The state of our Markov chain $\xi$ depends on multiplicities, coordinates and momenta of all particles: $\xi = \xi(N, \{N_s\}^{\mathrm{species}}, \{x_i\}_{i=1}^N, \{p_i\}_{i=1}^N)$. The initial state is an arbitrary set of particles that satisfy the required conservation laws (Eq. \ref{Eq:cons_laws}). Charge conservation for initial state is fulfilled by an \emph{ad hoc} heuristic algorithm picking lightest particles of necessary charges, while the energy-momentum conservation is achieved by rescaling momenta as in \cite{Schwarz:2017bdg}. Given a state $\xi$ we propose a state $\xi'$ with probability $T(\xi \to \xi')$ and then decide, if this state should be accepted, with probability $A(\xi \to \xi')$. Therefore, the probability to obtain a state $\xi'$ from $\xi$ is $w(\xi\to \xi') = T(\xi \to \xi') A(\xi \to \xi')$. The master equation, connecting the probability to obtain the state $\xi$ at steps $t$ and $t+1$ is

\begin{eqnarray}
P^{t+1}(\xi) - P^{t}(\xi) = \sum_{\xi'}  w(\xi' \to \xi) P^t(\xi') -\nonumber\\ w(\xi \to \xi') P^t(\xi) \,.
\end{eqnarray}

After many steps the probability $P^{t \to \infty}(\xi)$ should converge to $P(\xi)$ given by Eq. (\ref{Eq:sampled_distribution}). A sufficient condition for this is known as the detailed balance condition:

\begin{eqnarray}
\frac{P(\xi')}{P(\xi)} = \frac{w(\xi \to \xi')}{w(\xi' \to \xi)} =  \frac{T(\xi \to \xi')A(\xi \to \xi')}{T(\xi' \to \xi)A(\xi \to \xi')} \,.
\end{eqnarray}

This condition is satisfied if

\begin{eqnarray} \label{Eq:Metropolis_accept}
  a \equiv A(\xi \to \xi') = \textrm{min}\left(1, \, \frac{P(\xi') \, T(\xi' \to \xi)}{P(\xi) \, T(\xi \to \xi')}\right) \,.
\end{eqnarray}

 There is some freedom to select the proposal matrix $T(\xi\to\xi')$. We choose it such that it conserves energy, momentum, and quantum numbers. Consequently, our Markov chain never leaves the desired subspace where conservation laws are fulfilled. Our proposal matrix may be viewed as $2\to{3}$ and $3\to{2}$ stochastic ``collisions''~\cite{Seifert:2017oyb} on the hypersurface. However, we note, that there is no real time involved and ``collisions'' are not related to any physical process. They are simply a mathematical method  to sample the distribution of  Eq.~(\ref{Eq:sampled_distribution}). The proposal procedure is the following:
\begin{enumerate}
    \item With 50\% probability choose a $2\to 3$ or $3 \to 2$ transition.
    \item Select the ``incoming'' particles by uniformly picking one of all possible pairs or triples.
    \item Select the outgoing channel democratically with probability $1/N^{ch}$, where $N^{ch}$ is the number of possible channels, satisfying both quantum number and energy-momentum conservation.
    \item For the selected channel sample the ``collision'' kinematics uniformly from the available phase space with probability $\frac{dR_n}{R_n}$, $n = 2$ or $3$.
    \item Choose a cell for each of the outgoing particles uniformly from all cells in the patch. Note that this choice matters for the acceptance probability, because the corresponding temperatures, chemical potentials, velocities $u^{\mu}$, and normal 4-vectors $d\sigma_{\mu}$ in the Eq. (\ref{Eq:sampling_acceptance_probability}) will be taken at the cells, where the outgoing particles are thrown.
\end{enumerate}
Here $R_n$ is a phase-space integral for outgoing particles defined as the integral over $dR_n$:
\begin{eqnarray} \label{Eq:R_n_definition}
dR_n(\sqrt{s}, m_1, m_2, \dots, m_n) =  \frac{(2\pi)^4}{(2\pi)^{3n}}  \nonumber \\
\frac{d^3p_1}{2 E_1} \frac{d^3p_2}{2 E_2} \dots \frac{d^3p_n}{2 E_n} \delta^{(4)}(P_{tot}^{\mu} - \sum P_i^{\mu}) \,,
\end{eqnarray}
where $\sqrt{s} = (P_{tot}^{\mu} P^{tot}_{\mu})^{1/2}$. The integration of $dR_2$ and $dR_3$ is possible analytically \cite{Bauberger:1994nk,Seifert:2017oyb}. Our proposal procedure generates the following probabilities for $2\to 3$ and $3\to 2$ proposals:
\begin{eqnarray} \label{Eq:proposal_probabilities1}
T(2 \to 3) = \frac{1}{2} \frac{G_2^{ch}}{G_2} \frac{1}{N^{ch}_3} \frac{dR^{ch}_3}{R^{ch}_3} \frac{1}{N_{cells}^3}\\
\label{Eq:proposal_probabilities2}
T(3 \to 2) = \frac{1}{2} \frac{G_3^{ch}}{G_3} \frac{1}{N^{ch}_2} \frac{dR^{ch}_2}{R^{ch}_2} \frac{1}{N_{cells}^2}\,,
\end{eqnarray}
where $G_2 = \frac{N(N-1)}{2!}$ and $G_3 = \frac{N(N-1)(N-2)}{3!}$ denote total numbers of incoming pairs and triplets of any species, while $G_2^{ch}$ and $G_3^{ch}$ are the numbers of ways to select a given incoming particle species. Consequently, 
$\frac{G_2^{ch}}{G_2}$ and $\frac{G_3^{ch}}{G_3}$ represent the probabilities to obtain pairs and triplets of a given incoming species. The number of possible triplets and pairs of outgoing species with appropriate  quantum numbers are denoted by $N^{ch}_3$ and $N^{ch}_2$.
Inserting the proposal probabilities, Eqs.~(\ref{Eq:proposal_probabilities1}) and (\ref{Eq:proposal_probabilities2}), as well as the desired probability distribution, Eq.~(\ref{Eq:sampled_distribution}), into the expression for the acceptance probability,  Eq.~(\ref{Eq:Metropolis_accept}), we arrive, after some algebra, at  
\begin{eqnarray} \label{Eq:sampling_acceptance_probability}
a_{n\to m} = \frac{N^{ch}_m R_m}{N^{ch}_n R_n}  \frac{N!}{(N+m-n)!} \frac{m!}{n!} \frac{ k^{id}_m!}{ k^{id}_n!} \times \nonumber \\ \left( \frac{2 N_{cells}}{\hbar^3}\right)^{m-n} \frac{\displaystyle\prod_{i=1}^m g_i \, f_i(\mu_i - p_i^{\alpha}u_{\alpha},T) \, p^{\mu}_i d\sigma_{\mu}}{\displaystyle\prod_{j=1}^n g_j \, f_j(\mu_j - p^{\alpha}_j u_{\alpha},T) \, p^{\mu}_j d\sigma_{\mu}}
\end{eqnarray}
where we made use of the relation  $\prod \frac{d^3p_i}{(2\pi\hbar)^3 p^0_i} \delta^{(4)}(P_{tot}^{\mu} - \sum P_i^{\mu}) = 2^n \frac{dR_n}{(2\pi)^4} $. Here $n=2,3$ and $m=3,2$ are the numbers of incoming and outgoing particles, and $N$ is the total number of particles before proposing the Markov chain step. The product in the numerator is taken over the outgoing particles and the one in the denominator is taken over the incoming particles. The quantities $d\sigma$, $u$, $T$, $\mu$ should be evaluated in the cell where the particles are proposed to be, or coming from. The  total number of particles in the entire patch is given by $N$, and $k^{id}_m$ and $k^{id}_n$ are the numbers of outgoing and incoming identical species in the reaction. Note that the sampling accounts for  the variations in temperature and chemical potential within the patch. Also, and equally important, the distribution function $f$ may contain viscous corrections. To summarize, the algorithm consists of multiple Markov chain steps, where the step is proposed according to Eqs.~(\ref{Eq:proposal_probabilities1}) and (\ref{Eq:proposal_probabilities2}) and accepted with probability given by Eq.~(\ref{Eq:sampling_acceptance_probability}).

\begin{figure*}[h!tb]
    \centering
    \includegraphics[height=5.8cm]{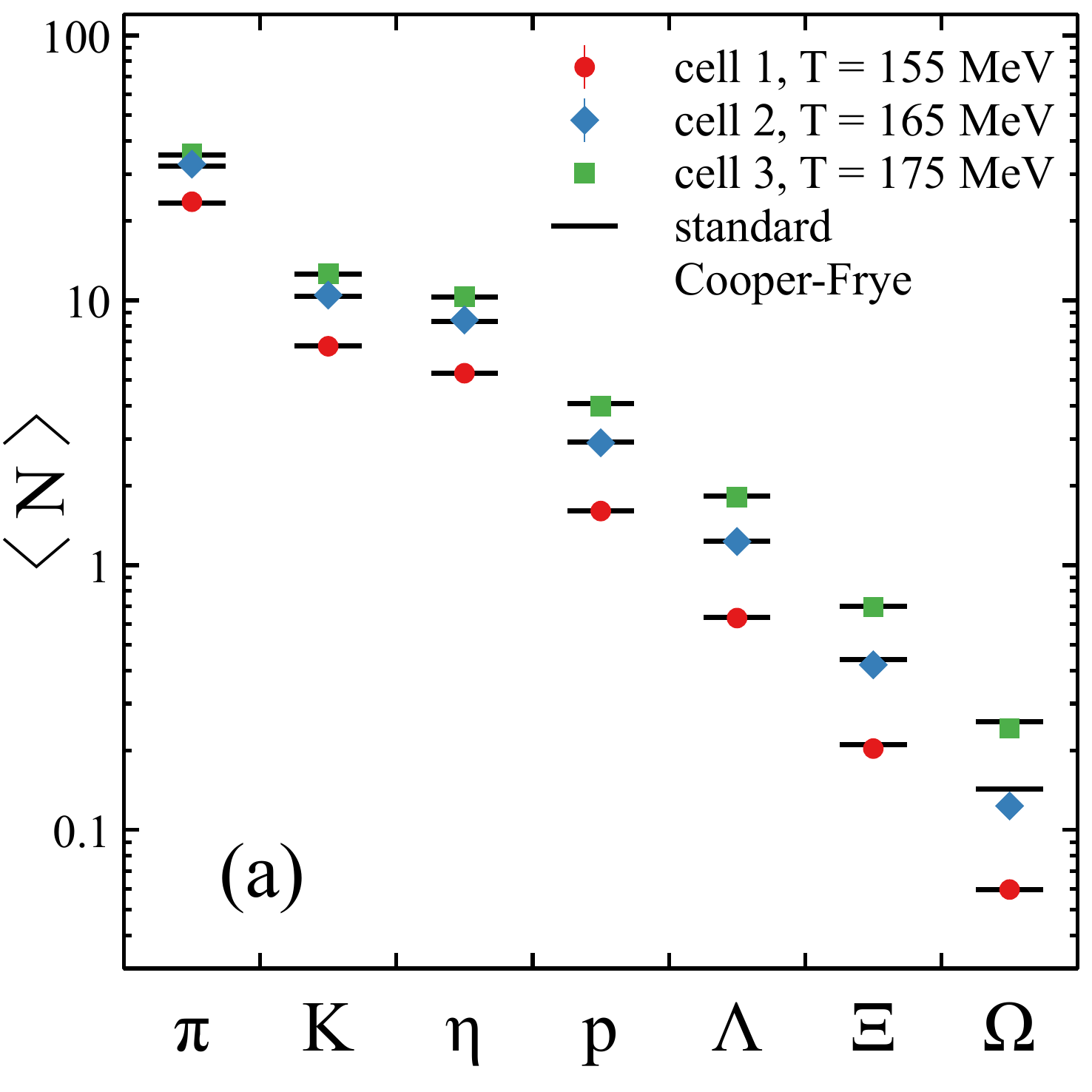}
    \includegraphics[height=5.8cm]{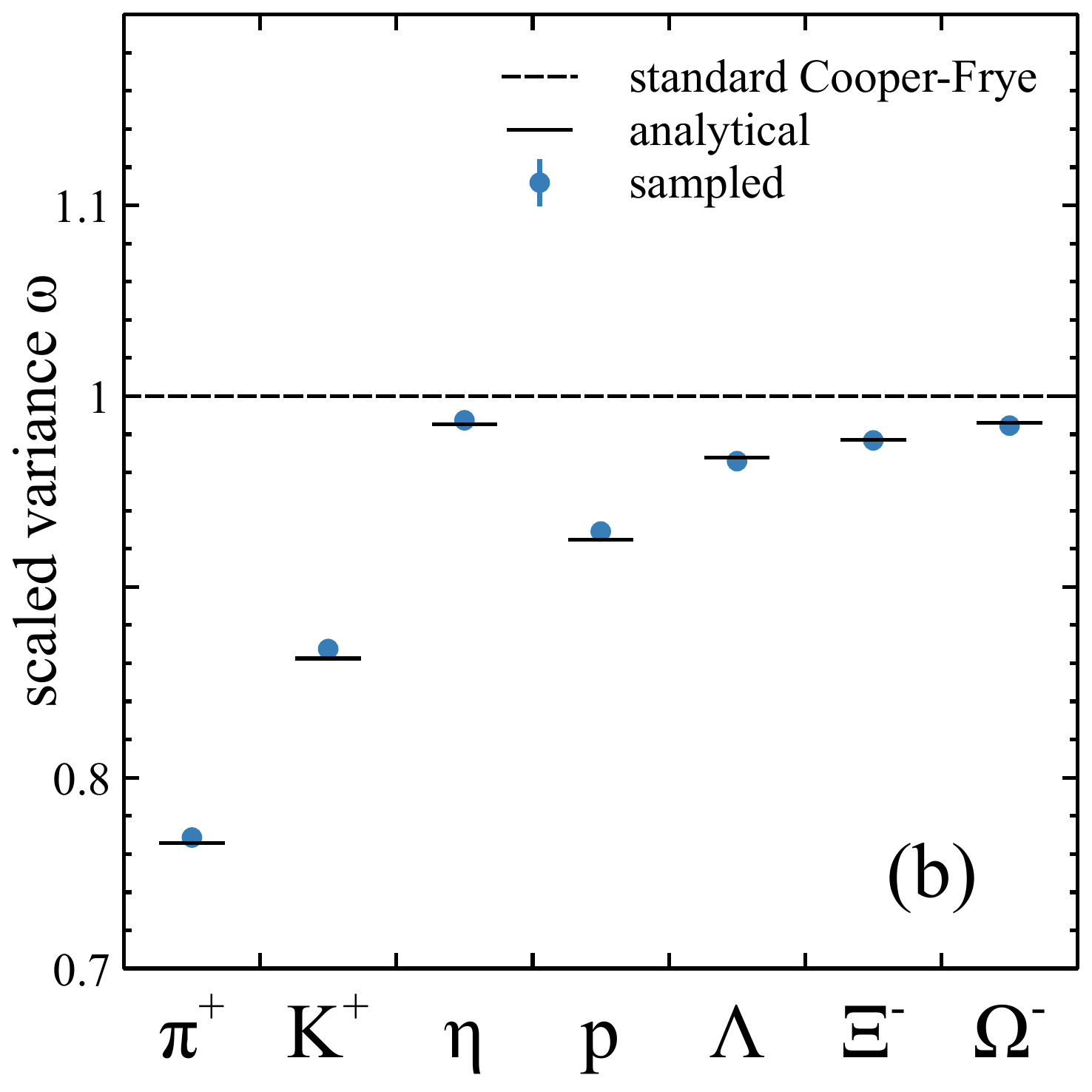}
    \includegraphics[height=5.8cm]{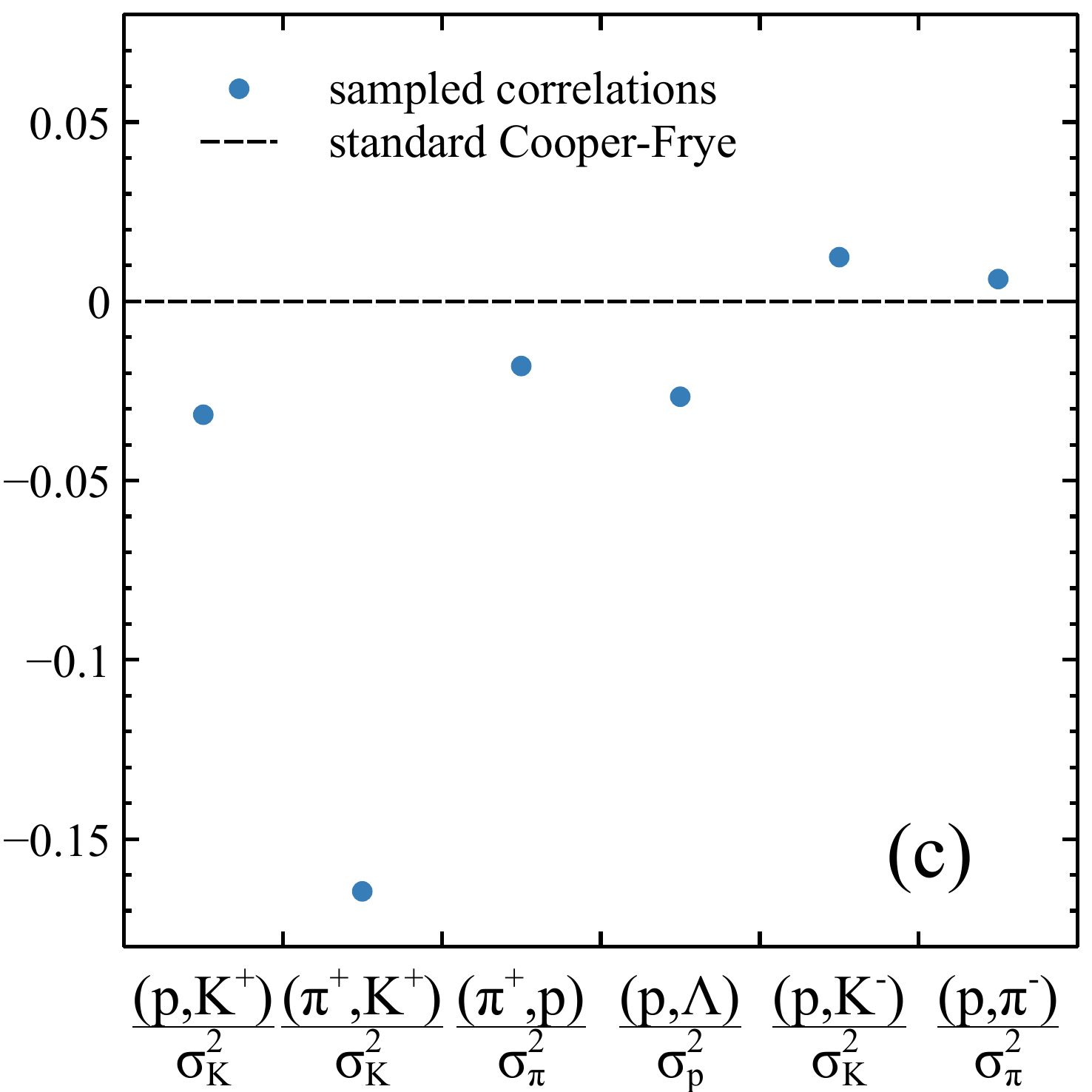}
    \caption{
      Demonstration of the sampling with conservation laws over the patch, where total baryon number, strangeness, and charge are enforced to be 0, while total energy and momentum are fixed and given by Eq.~(\ref{Eq:cons_laws}). The patch consists of 3 cells with arbitrarily selected normals $d\sigma^{\mu}_1 = (500.0, 50.0, 20.0, 30.0)\,\rm fm^{3}$, $d\sigma^{\mu}_2 = (500.0, 40.0, 80.0, 30.0) \,\rm fm^{3}$, $d\sigma^{\mu}_3 = (500.0, 20.0, 20.0, 20.0) \,\rm fm^{3}$; collective velocities $\vec{v}_1 = (0.2, 0.3, 0.4)$, $\vec{v}_2 = (0.1, 0.5, 0.5)$, $\vec{v}_3 = (0.3, 0.4, 0.2)$; and temperatures $T_1 = 0.155$ GeV, $T_2 = 0.165$ GeV, $T_3 = 0.175$ GeV. Mean multiplicities of selected hadrons in the cells are shown in panel (a): they are unchanged compared to standard grand-canonical Cooper-Frye sampling. However, the scaled variances of multiplicities in the whole patch, shown in panel (b), differ from the standard Cooper-Frye result and coincide within 0.5\% with the microcanonical expectation in the thermodynamic limit, computed using analytic formulas from~\cite{Hauer:2007ju}. In panel (c), the non-trivial correlations, generated by conservation laws, are shown in contrast to no correlations in the standard Cooper-Frye sampling. Correlations are defined as $(A,B) \equiv \langle (A - \langle A \rangle) (B - \langle B\rangle) \rangle$, where $\langle \rangle$ denotes average over samples; $\sigma_A^2 \equiv (A,A)$. }
    \label{Fig:I}
\end{figure*}
We have tested the  above sampling algorithm both in a simplified and a somewhat more realistic setup. First, we consider a patch consisting of one cell with $u^{\mu} = (1,0,0,0)$, $d\sigma_{\mu} = (V, 0,0,0)$, and $f(p) = \exp(-p^0/T)$. In this case Eq.~(\ref{Eq:sampled_distribution}) represents a well-known microcanonical distribution, which has been sampled before~\cite{Werner:1995mx,Becattini:2004rq}. Additionally, analytical expectations of scaled variances of hadron multiplicities in the large volume limit are available in this case~\cite{Hauer:2007ju}. Our algorithm reproduces the analytical results for both means and scaled variances well. The resulting momentum distributions are very close to Boltzmann, as expected. This is a non-trivial result, because multiplicity and momentum distributions are a not a direct input to the sampling; the only input is volume $V$, total energy, momentum and conserved charges.

Next, we demonstrate our sampling for a more realistic scenario, where we consider a patch consisting of three cells with non-trivial  values  for $u^{\mu}$, $d\sigma_{\mu}$, and $T$, which also vary from cell to cell
\footnote{We have also successfully tested the above algorithm including patch selection on realistic hyper-surfaces, which will be reported on in detail in a subsequent publication.}
. Conservation laws are imposed over the entire patch, while the local energy density and charge densities vary from cell to cell. As shown in Fig.~\ref{Fig:I}(a), we obtain the expected means in each cell, agreeing with the grand-canonical standard Cooper-Frye sampling. This is a demonstration that we correctly reproduce the temperature variation from cell to cell. The scaled variances, $\omega$, of multiplicities in the patch, shown in Fig.~\ref{Fig:I}(b), are already drastically different from the standard grand-canonical Cooper-Frye sampling, where $\omega = 1$ by construction. In our case the variances agree with the microcanonical analytical expectation from~\cite{Hauer:2007ju}. Finally, in Fig.~\ref{Fig:I}(c) we demonstrate the non-trivial correlations emerging from conservation laws. Unlike for variances, to our knowledge, there is no analytical calculation of correlations in a microcanonical ensemble, although in principle such calculations are possible using techniques developed in~\cite{Hauer:2007ju}. Beside testing the sampler implementation, Fig.1~\ref{Fig:I}(b,c) also shows the expected effect of conservation laws on fluctuations and correlations in heavy ion collisions.

In summary, we have proposed, implemented and tested a particlization method which takes into account local event-by-event conservation laws in a systematic fashion. Localness is achieved by splitting the hypersurface into patches and enforcing conservation laws in every patch. Event-by-event conservation of total energy, momentum, baryon number, strangeness, and electric charge over the patch is guaranteed by the algorithm. At the same time local cell-by-cell variations of energy and charge densities within a patch are preserved, ensuring that observables sensitive to these variations, such as higher order azimuthal asymmetries~\cite{Gardim:2011xv}, are not smeared out. The proposed method is essential for studies of correlations and fluctuations, especially in combination with stochastic hydrodynamics, since it will not obfuscate its correlations and fluctuations. It may also be applied to exploring small systems, where the impact of event-by-event conservation laws is large, as well as charge dependent correlations relevant for the the chiral magnetic effect \cite{Voloshin:2004vk}. The code used for sampling  is publicly available~\cite{Oliinychenko:code}. We have checked that its execution time for realistic hypersurfaces is not impractical \footnote{The overall runtime to generate $10^5$ samples for our  test scenario constituted around 2 minutes. For a realistic hypersurface expected from a heavy ion collision of $\sqrt{s}\simeq 10 \,\rm GeV$. where we have about 150 patches of $\sim 10\,\rm GeV$, the present implementation takes about 40 minutes to generate 10000 samples} As a next step, we will apply it to search for observable effects of critical fluctuations in heavy ion collisions. Through accounting for local conservation laws we might be able to detect critical fluctuations in coordinate space via correlations in momentum space which were previously not visible when using standard particlization.

\begin{acknowledgments}
We would like to thank members of the BEST collaboration for fruitful discussions, and
A.~Wergieluk for a critical reading of the manuscript.
This work was supported by the U.S. Department of Energy, 
Office of Science, Office of Nuclear Physics, under contract number 
DE-AC02-05CH11231 and received support within the framework of the
Beam Energy Scan Theory (BEST) Topical Collaboration.
\end{acknowledgments}


\begin{thebibliography}{99}

\bibitem{LL_T5_par146}
  L.~D.~Landau and E.~M.~Lifshitz, T. 5 "Statistical Physics", Part 1, 3rd edition, Moscow 1972,
  §146.
  

\bibitem{Stephanov:2008qz} 
   M.~A.~Stephanov,
   Phys.\ Rev.\ Lett.\  {\bf 102}, 032301 (2009)
   doi:10.1103/PhysRevLett.102.032301
   [arXiv:0809.3450 [hep-ph]].


\bibitem{Luo:2015ewa} 
  X.~Luo [STAR Collaboration],
  PoS CPOD {\bf 2014}, 019 (2015)
  doi:10.22323/1.217.0019
  [arXiv:1503.02558 [nucl-ex]].
  
\bibitem{Adamczyk:2017wsl}
  L.~Adamczyk {\it et al.} [STAR Collaboration],
  Phys.\ Lett.\ B {\bf 785} (2018) 551
  doi:10.1016/j.physletb.2018.07.066
  [arXiv:1709.00773 [nucl-ex]].
  
\bibitem{Abdelwahab:2014yha} 
  N.~M.~Abdelwahab {\it et al.} [STAR Collaboration],
  Phys.\ Rev.\ C {\bf 92}, no. 2, 021901 (2015)
  doi:10.1103/PhysRevC.92.021901
  [arXiv:1410.5375 [nucl-ex]].
  
\bibitem{Adam:2019rsf} 
  J.~Adam {\it et al.} [STAR Collaboration],
  arXiv:1901.00837 [nucl-ex].
  
\bibitem{Adamczyk:2015yga} 
  L.~Adamczyk {\it et al.} [STAR Collaboration],
  Phys.\ Rev.\ C {\bf 94}, no. 2, 024909 (2016)
  doi:10.1103/PhysRevC.94.024909
  [arXiv:1507.03539 [nucl-ex]].
  
\bibitem{Hirano:2012kj} 
  T.~Hirano, P.~Huovinen, K.~Murase and Y.~Nara,
  Prog.\ Part.\ Nucl.\ Phys.\  {\bf 70}, 108 (2013)
  doi:10.1016/j.ppnp.2013.02.002
  [arXiv:1204.5814 [nucl-th]].
  
\bibitem{Ryu:2017qzn} 
  S.~Ryu, J.~F.~Paquet, C.~Shen, G.~Denicol, B.~Schenke, S.~Jeon and C.~Gale,
  Phys.\ Rev.\ C {\bf 97}, no. 3, 034910 (2018)
  doi:10.1103/PhysRevC.97.034910
  [arXiv:1704.04216 [nucl-th]].
  
\bibitem{Ahmad:2016ods} 
  S.~Ahmad, H.~Holopainen and P.~Huovinen,
  Phys.\ Rev.\ C {\bf 95}, no. 5, 054911 (2017)
  doi:10.1103/PhysRevC.95.054911
  [arXiv:1608.03444 [nucl-th]].
  
\bibitem{Donev:2009}
  A.~Donev, J.~B.~Bell, A.~L.~Garcia and B.~J.~Alder,
  Multiscale\ Model.\ Simul., 8(3), 871–911
  doi:10.1137/090774501
  [arXiv0910.3968 [cond-mat.soft]]
  
\bibitem{Kapusta:2011gt} 
  J.~I.~Kapusta, B.~Muller and M.~Stephanov,
  Phys.\ Rev.\ C {\bf 85}, 054906 (2012)
  doi:10.1103/PhysRevC.85.054906
  [arXiv:1112.6405 [nucl-th]].

\bibitem{Murase:2013tma} 
  K.~Murase and T.~Hirano,
  arXiv:1304.3243 [nucl-th].

\bibitem{Kumar:2013twa} 
  A.~Kumar, J.~R.~Bhatt and A.~P.~Mishra,
  Nucl.\ Phys.\ A {\bf 925}, 199 (2014)
  doi:10.1016/j.nuclphysa.2014.02.012
  [arXiv:1304.1873 [hep-ph]].
  
\bibitem{Nahrgang:2018afz} 
  M.~Nahrgang, M.~Bluhm, T.~Schäfer and S.~A.~Bass,
  arXiv:1804.05728 [nucl-th].
  
\bibitem{Nahrgang:2011mg} 
  M.~Nahrgang, S.~Leupold, C.~Herold and M.~Bleicher,
  Phys.\ Rev.\ C {\bf 84}, 024912 (2011)
  doi:10.1103/PhysRevC.84.024912
  [arXiv:1105.0622 [nucl-th]].
  
\bibitem{Herold:2014zoa} 
  C.~Herold, M.~Nahrgang, Y.~Yan and C.~Kobdaj,
  J.\ Phys.\ G {\bf 41}, no. 11, 115106 (2014)
  doi:10.1088/0954-3899/41/11/115106
  [arXiv:1407.8277 [hep-ph]].
  
\bibitem{Cooper:1974mv} 
  F.~Cooper and G.~Frye,
  Phys.\ Rev.\ D {\bf 10}, 186 (1974).
  doi:10.1103/PhysRevD.10.186
  
\bibitem{Bozek:2012en} 
  P.~Bozek and W.~Broniowski,
  Phys.\ Rev.\ Lett.\  {\bf 109}, 062301 (2012)
  doi:10.1103/PhysRevLett.109.062301
  [arXiv:1204.3580 [nucl-th]].
  
\bibitem{Huovinen:2012is} 
  P.~Huovinen and H.~Petersen,
  Eur.\ Phys.\ J.\ A {\bf 48}, 171 (2012)
  doi:10.1140/epja/i2012-12171-9
  [arXiv:1206.3371 [nucl-th]].

\bibitem{Schwarz:2017bdg} 
  C.~Schwarz, D.~Oliinychenko, L.-G.~Pang, S.~Ryu and H.~Petersen,
  J.\ Phys.\ G {\bf 45}, no. 1, 015001 (2018)
  doi:10.1088/1361-6471/aa90eb
  [arXiv:1707.07026 [hep-ph]].
  
\bibitem{Oliinychenko:2016vkg} 
  D.~Oliinychenko and H.~Petersen,
  J.\ Phys.\ G {\bf 44}, no. 3, 034001 (2017)
  doi:10.1088/1361-6471/aa528c
  [arXiv:1609.01087 [nucl-th]].
  
\bibitem{An:2019osr} 
  X.~An, G.~Basar, M.~Stephanov and H.~U.~Yee,
  arXiv:1902.09517 [hep-th].

\bibitem{Steinheimer:2017dpb} 
  J.~Steinheimer and V.~Koch,
  Phys.\ Rev.\ C {\bf 96}, no. 3, 034907 (2017)
  doi:10.1103/PhysRevC.96.034907
  [arXiv:1705.08538 [nucl-th]].
  
\bibitem{Becattini:2004rq} 
  F.~Becattini and L.~Ferroni,
  Eur.\ Phys.\ J.\ C {\bf 38}, 225 (2004)
  Erratum: [Eur.\ Phys.\ J.\  {\bf 66}, 341 (2010)]
  doi:10.1140/epjc/s10052-010-1243-4, 10.1140/epjc/s2004-02027-8
  [hep-ph/0407117].

\bibitem{Werner:1995mx} 
  K.~Werner and J.~Aichelin,
  Phys.\ Rev.\ C {\bf 52}, 1584 (1995)
  doi:10.1103/PhysRevC.52.1584
  [nucl-th/9503021].
  
\bibitem{Seifert:2017oyb} 
  E.~Seifert and W.~Cassing,
  Phys.\ Rev.\ C {\bf 97}, no. 2, 024913 (2018)
  doi:10.1103/PhysRevC.97.024913
  [arXiv:1710.00665 [hep-ph]].
  
\bibitem{Bauberger:1994nk} 
  S.~Bauberger, M.~B\"ohm, G.~Weiglein, F.~A.~Berends and M.~Buza,
  Nucl.\ Phys.\ Proc.\ Suppl.\  {\bf 37B}, no. 2, 95 (1994)
  doi:10.1016/0920-5632(94)90665-3
  [hep-ph/9406404].
  
\bibitem{Hauer:2007ju} 
  M.~Hauer, V.~V.~Begun and M.~I.~Gorenstein,
  Eur.\ Phys.\ J.\ C {\bf 58}, 83 (2008)
  doi:10.1140/epjc/s10052-008-0724-1
  [arXiv:0706.3290 [nucl-th]].

\bibitem{Gardim:2011xv} 
  F.~G.~Gardim, F.~Grassi, M.~Luzum and J.~Y.~Ollitrault,
  Phys.\ Rev.\ C {\bf 85}, 024908 (2012)
  doi:10.1103/PhysRevC.85.024908
  [arXiv:1111.6538 [nucl-th]].
  
  \bibitem{Voloshin:2004vk} 
  S.~A.~Voloshin,
  Phys.\ Rev.\ C {\bf 70}, 057901 (2004)
  doi:10.1103/PhysRevC.70.057901
  [hep-ph/0406311].
  
\bibitem{Oliinychenko:code}
  \href{}{github.com/doliinychenko/microcanonical_cooper_frye}
  

\end{thebibliography}
\end{document}